\documentclass[11pt]{article}
\usepackage{natbib}
\usepackage{amsmath}
\usepackage{color}
\usepackage{latexsym}
\usepackage{epsfig}
\usepackage{float,verbatim}

\newcommand{\xVec}{{\bf x}}
\newcommand{\xnew}{\xVec^{\rm new}}
\newcommand{\ymaxn}[1]{{y_{\max}^{(#1)}}}
\newcommand{\yminn}[1]{{y_{\min}^{(#1)}}}

\begin{document}

%\title{Overview of Expected Improvement - based Sequential Designs in Computer Experiments}
\title{Design of Computer Experiments for Optimization, Estimation of Function Contours,
and Related Objectives}

\date{}

\maketitle

\begin{center}
Derek Bingham, Pritam Ranjan, and William J.\ Welch \\
Simon Fraser University, Acadia University, and University of British Columbia
\end{center}

\section{Introduction}\label{sect:intro}

A computer code or simulator is a mathematical representation of a physical system,
for example a set of differential equations.
% JL
Such simulators take a set of input values or conditions, $\xVec$, 
and from them produce an output value, $y(\xVec)$, or several such outputs. 
%Running the code with given values of the vector of inputs, $\xVec$,
%leads to an output $y(\xVec)$ or several such outputs.
For instance, one application we use for illustration simulates
the average tidal power, $y$, generated as a function of a turbine location,
$\xVec = (x_1, x_2)$,
in the Bay of Fundy, Nova Scotia, Canada
\citep{RanHayKar2011}.
Performing scientific or engineering experiments via such a computer code
is often more time and cost effective than running a physical experiment
% JL  
or collecting data directly.
%Or a physical experiment might be infeasible.

%A computer experiment is a designed set of runs of the computer code
%Advantages.
%Deterministic.
%\cite{SacWelMit1989}, \cite{CurMitMor1991}, and \cite{Oha1992}

A computer experiment often has similar objectives to a physical experiment.
% JL
For example, computer experiments are often used in manufacturing or process development. 
If $y$ is a quality measure for a product or process,
an experiment could aim to optimize $y$ with respect to $\xVec$.
Similarly, an experiment might aim to find sets or contours of $\xVec$ values
that make $y$ equal a specified target value.
Such scientific and engineering objectives are naturally and efficiently achieved
via so-called data-adaptive sequential design,
which we describe below.
Essentially, each new run (that is, new set of input values) is chosen 
based on analysis of the data so far,
to make the best expected improvement in the objective. 
In a computer experiment,
choosing new experimental runs, re-starting the experiment, etc.\ pose
only minor logistical challenges if these decisions are also computer-controlled,
a distinct advantage relative to a physical experiment.

Choosing new runs sequentially for optimization, moving $y$ to a target, etc.\ has
been formalized using the concept of expected improvement \citep{Jones1998}.
The next experimental run is made where the expected improvement in the function of interest
%(e.g., the maximum of the function) has largest expectation.
is largest.
This expectation is with respect to the predictive distribution of $y$ from
a statistical model relating $y$ to $\xVec$.
By considering a set of possible inputs $\xVec$ for the new run, 
we can choose that which gives the largest expectation.

We illustrate this basic idea with two examples in Section~\ref{sect:basic}.
Then we describe formulations of improvement functions and their expectations
in Section~\ref{sect:EI}.
Expectation implies a statistical model,
and in Section~\ref{sect:GP} we outline the use of Gaussian process models
for fast emulation of computer codes.
In Section~\ref{sect:other} we describe some extensions to other, more complex scientific
objectives.

\section{Expected improvement and sequential design: basic ideas \label{sect:basic}}

We illustrate the  basic idea of expected improvement and data-adaptive sequential
design via two examples.
The first,
a tidal-power application, % described by \cite{RanHayKar2011},
shows the use of expected improvement in sequential optimization.
We then use a simulator of volcanic pyroclastic flow
to illustrate how to map out a contour of a function.

\subsection{Optimization}

\cite{RanHayKar2011} described output from a
2D computer-model simulation of the power produced by
a tidal turbine in the Minas Passage of the Bay of Fundy,
Nova Scotia, Canada.
In this simplified version of the problem there are just two inputs 
for the location of a turbine.
Originally, the input space was defined by latitude-longitude coordinates
for a rectangular region in the Minas Passage 
\citep[see Figure~5 of][]{RanHayKar2011}. 
The coordinates  were transformed so that 
$x_1$ is in the direction of the flow  
and $x_2$ is perpendicular to the flow.
%(i.e., axes of the colored region in Figure 5).
Furthermore, only an interesting part of the Minas Passage was considered,
with $x_1 \in [0.75, 0.95]$ and $x_2 \in [0.2, 0.8]$. 
The code generates $y$, the extractable power in MW,
averaged over a tidal cycle.

\begin{figure}
\centerline{\epsfig{file=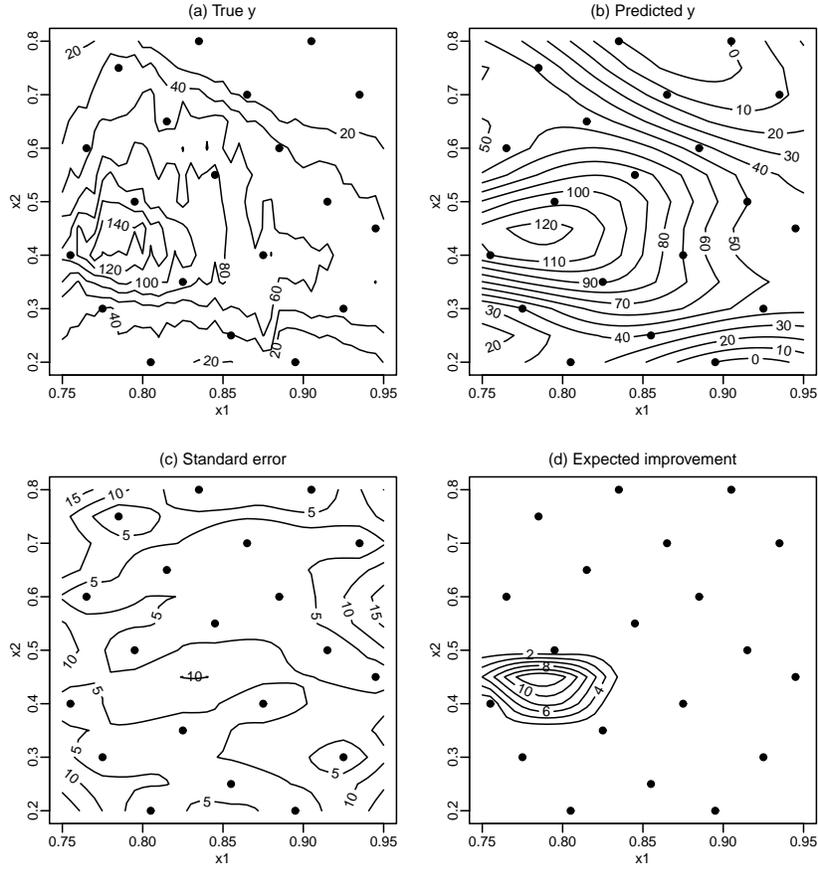,width=4.4in}}
\vspace{-0.5cm}
\caption{Initial 20-run design and analysis for the tidal-power application:
(a) true power, $y$, in MW;
(b) predicted power, $\hat{y}(\xVec)$;
(c) standard error, $s(\xVec)$; and
(d) expected improvement, $E[I(\xVec)]$.
The design points from an initial 20-run maximin Latin hypercube are
shown as filled circles.
All plots are functions of the two input variables, $x_1$ and $x_2$,
which are transformations of longitude and latitude.
%In (d) only the part of the input space with substantial values of EI is shown.
\label{fig:tidal:ei20}}
\end{figure}

For the simplified demonstration here,
$y$ was computed for 533 runs on a $13 \times 41$ grid of $x_1$ and $x_2$ values,
which produced the contour plot of Figure~\ref{fig:tidal:ei20}(a).

We now demonstrate how the turbine location optimizing the power,
i.e., $\max y(x_1, x_2)$, can be found with far fewer than 533 runs of
the computer code.
Such an approach would be essential for the computer experiment of ultimate interest.
A more realistic computer model has a grid resolution 10 times finer
in each coordinate and introduces vertical layers in a 3D code.
The running time would be increased by several orders of magnitude.
Moreover, the final aim is to position several turbines,
which would interfere with each other,
and so the optimization space is larger than two or three dimensions.
Thus, the ultimate goal is to optimize a high-dimensional function
with a limited number of expensive computer model runs.
Inevitably, much of the input space cannot be explicitly explored,
and a statistical approach to predict outcomes (extractable power) 
along with an uncertainty measure
is required to decide where to make runs and when to stop.
The expected improvement criterion addresses these two requirements.

%PRITAM: Can you explain how the longitude-latitude coordinates in your paper's
%Figure 5 become the $x_1, x_2$ coordinates you gave me?  Just some sort of rotation
%of the rectangle of interest?)
%\textcolor{red}{Will: Richard Karsten (my coauthor) drew Figure~5 and generated the data for Figure~6.
%When we were writing the paper, somehow I did not think about the transformation of the axis (both
%looked pretty rectangular so I assumed that the transformation was easy). Of course, I should have
%changed the axes to $[0, 1]^2$ (I should have noticed it then). Apparently, Richard did some non-
%trivial mapping (numerically) based on the geography of Minas Passage and direction of flow.
%Unfortunately,  there is no simple formula that he can provide for this transformation.})

Thus, imagine a more limited computer experiment with just 20 runs,
as shown by the points in Figure~\ref{fig:tidal:ei20}.
The experimental design  
% JL
(that is, the locations of the 20 points)
is a maximin Latin hypercube \citep{MorMit1995},
a stratified scheme that is ``space-filling''
even in higher dimensions.
The choice of 20 runs is based on the heuristic rule that
an initial computer experiment has $n = 10d$ observations \citep{LoeSacWel2009},
where $d$ is the input dimension; here $d = 2$.
Among the 20 initial runs, the largest $y$ observed, denoted by $\ymaxn{20}$,
is 109.7 MW at $(x_1, x_2) = (0.755, 0.4)$.
The expected improvement algorithm tries to improve on the best value
found so far as new runs are added.

At each iteration of the computer experiment we 
obtain a predictive distribution for $y(\xVec)$ conditional on the runs so far.
This allows prediction of the function at input vectors $\xVec$ where
the code has not been run.
%In the computer experiment literature,
A Gaussian process (GP) statistical model is commonly used for prediction,
as outlined in Section~\ref{sect:GP},
though this is not essential.
A GP model was fit here to the data from the first 20 runs, giving
the point-wise predictions, $\hat{y}(\xVec)$, of $y(\xVec)$
in Figure~\ref{fig:tidal:ei20}(b)
and the standard error, $s(\xVec)$,
in Figure~\ref{fig:tidal:ei20}(c).
The standard error is a statistical measure of uncertainty concerning the closeness of the predicted value to the actual true value of y(x).  We show the predicted values and standard errors through contours in Figures~\ref{fig:tidal:ei20}(b) and~\ref{fig:tidal:ei20}(c).

Figures~\ref{fig:tidal:ei20}(b) and~(c) are informative about
regions in the input space that are promising versus unpromising
for further runs of the code.
While the $\hat{y}(\xVec)$ prediction surface is nonlinear,
it suggests there is a single, global optimum.
Moreover, the $s(\xVec)$ surface is uniformly
below about 15:
For much of the input space, $\hat{y}(\xVec)$ is so much smaller than
$\ymaxn{20}$ relative to $s(\xVec)$ 
that a new run is expected to make virtually zero improvement.

The expected improvement (EI) for a candidate new run at any $\xVec$
is computed from the predictive distribution of $y(\xVec)$.
(See Section~\ref{sect:EI:opt} for the formal definition of EI.)
Figure~\ref{fig:tidal:ei20}(d) shows the EI surface based
on predictive distributions from a GP that is fitted to the  data from 
the initial 20 runs of the tidal-power code.
By the definition in Section~\ref{sect:EI:opt},
improvement can never be negative
(if the output from the new run does beat the current optimum, the current optimum stands).
Thus, EI is always non-negative too.
Figure~\ref{fig:tidal:ei20}(d) indicates that for most of the input space 
EI is near zero and a new run would be wasted,
but there is a sub-region where EI is more than 12 MW.
Evaluating EI over the $13 \times 41$ grid shows that the maximum EI is
13.9 at $\xVec = (0.785, 0.45)$.
In other words, a new code run to evaluate $y(0.785, 0.45)$ 
is expected to beat $\ymaxn{20} = 109.7$ by about 14.

Thus, run 21 of the sequential design for the computer experiment is at 
$\xVec^{(21)} = (0.785, 0.45)$.
The actual power obtained from the simulator is $y = 159.7$ MW,
so the best $y$ found after 21 runs is $\ymaxn{21} = 159.7$,
and this is the value to beat at the next iteration.
Note that the actual improvement in the optimum from the new run is $159.7 - 109.7 = 50.0$,
compared with an expectation of about 13.9.

The new run raises concerns about the statistical model. 
Before making the new run, 
the predictive distribution of $y(\xVec^{(21)})$ 
is approximately normal, $N(123.5, 5.67^2)$,
an implausible distribution given the large value of the standardized residual
$(y(\xVec^{(21)}) - \hat{y}(\xVec^{(21)})) / s(\xVec^{(21)}) = (159.7 - 123.5) / 5.67 = 6.4$.
One deficiency is that $s(\xVec)$ may not reflect all sources of uncertainty in
estimation of the parameters of the GP (see Section~\ref{sect:GP}).
A more important reason here, however, is that the new observation successfully 
finds a peak in the input space, 
a sub-region where the output function is growing rapidly and uncertainty is larger.
In contrast, the first 20 runs were at locations where the function is flatter
and easier to model.
The GP model fit to the initial runs under-estimated the uncertainty of prediction 
in a more difficult part of the input space.

Careful consideration of the properties of a GP model and the possible
need for transformations is particularly relevant
for sequential methods based on predictive distributions.
Uncertainty of prediction is a key component of the EI methodology,
so checking that a model has plausible standard errors of prediction
is critical.

One way of improving the statistical emulator of the tidal-power code
is to consider transformation of the output.
This is described in the context 
of the volcano example of Section~\ref{sect:basic:contour},
where transformation is essential.
For the tidal-power example, 
persisting with the original model will show that it adapts to give
more plausible standard errors with a few more runs.

\begin{figure}
\centerline{\epsfig{file=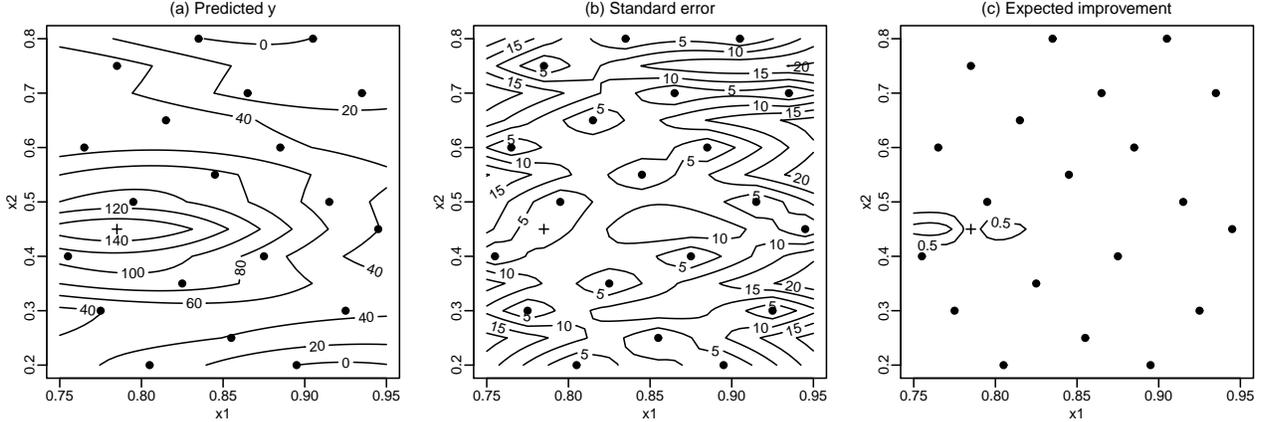,width=6.7in}}
\vspace{-0.5cm}
\caption{Analysis of the tidal-power application after 21 runs:
(a) predicted power, $\hat{y}$;
(b) standard error, $s(\xVec)$; and
(c) expected improvement, $I(\xVec)]$.
The new design point is shown as a ``$+$''.
\label{fig:tidal:ei21}}
\end{figure}

The GP model and predictive distributions are next updated
to use the data from all 21 runs now available.
%This happens to be the maximum of $y$ over the grid.
%Based on the runs so far, though, we do not know yet whether we are done,
%and the algorithm iterates.
Figure~\ref{fig:tidal:ei21}(a) shows the location of the new run as a ``$+$''
and the updated $\hat{y}(\xVec)$.
Similarly, Figure~\ref{fig:tidal:ei21}(b) gives the updated $s(\xVec)$.
A property of the GP fit is that $s(\xVec)$ must be zero at any point $\xVec$
where $y(\xVec)$ is in the data set for the fit
(see \cite{Jones1998} for a derivation of this result).
Thus, $s(\xVec)$ is zero at the new run, 
and Figure~\ref{fig:tidal:ei21}(b) shows it is less than 5 near the new run.
Comparing with Figure~\ref{fig:tidal:ei20}(c),
it is seen that $s(\xVec)$ was 5 or more in this neighbourhood for the GP
fit before the new run.

On the other hand, comparison of Figures~\ref{fig:tidal:ei20}(c) and~\ref{fig:tidal:ei21}(b)
shows that $s(\xVec)$ has {\em increased\/} outside the neighbourhood of the new run.
For example, at the right edge of Figure~\ref{fig:tidal:ei20}(c),
$s(\xVec)$ barely reaches 15,
yet $s(\xVec)$ often exceeds 15 or even 20 at the same locations in Figure~\ref{fig:tidal:ei21}(b).
The 21-run GP fit has adapted to reflect the observed greater sensitivity of the output
to $x_1$ and $x_2$.
(For instance, the estimate of the GP variance parameter $\sigma^2$, 
defined in Section~\ref{sect:GP}, increases.)
Thus, the model has at least partially self corrected and we continue with it.

The EI contour plot in Figure~\ref{fig:tidal:ei21}(c)
suggests that there is little further improvement to be had from a further run anywhere.
If a run number 22 is made, however,
it is not located where $\hat{y}(\xVec)$ is maximized;
that location coincides with run 21 and there would be no gain.
Rather, the maximum EI of about 1.3 MW occurs
at a moderate distance from the location providing maximum $\hat{y}(\xVec)$.
As we move away from run 21, the standard error increases from zero
until it is large enough to allow a modest expected improvement.
Thus, this iteration illustrates that EI trades off local search
(evaluate where $\hat{y}(\xVec)$ is optimized) and global search
(evaluate where uncertainty concerning fitted versus actual output values, 
characterized by $s(\xVec)$, is optimized).

With this approach, EI typically indicates
smaller potential gains as the number of iterations increases.
Eventually, the best EI is deemed small enough to stop.
It turns out that run 21 found the global maximum for extractable
power on the $13 \times 41$ grid of locations.

\subsection{Contour estimation}\label{sect:basic:contour}

We illustrate sequential design for mapping out a contour of a
computer-model function using TITAN2D computer model runs
provided by Elaine Spiller.
They relate to the Colima volcano in Mexico.
%a particularly hazardous volcano.
Again for ease of illustration, there are two input variables:
$x_1$ is the pyroclastic flow volume ($\mbox{m}^3$) of fluidized gas and rock fragments
from the eruption;
and $x_2$ is the basal friction angle in degrees,
defined as the the minimum slope for the volcanic material to slide.
The output $z$ is the maximum flow height (m) at a single,
critical location.
As is often the case, the code produces functional output,
here flow heights over a 2D grid on the earth's surface,
but the output for each run is reduced to a scalar quantity of interest,
the height at the critical location.

Following \cite{BayBerCal2009},
the scientific objective is to find the values of $x_1$ and $x_2$ where
$z=1$, a contour delimiting a ``catastrophic'' region.
\cite{BayBerCal2009} used the same TITAN2D code but for a different volcano.
They also conducted their sequential experiment in a less
formal way than in our illustration of the use of EI.

\begin{figure}
\centerline{\epsfig{file=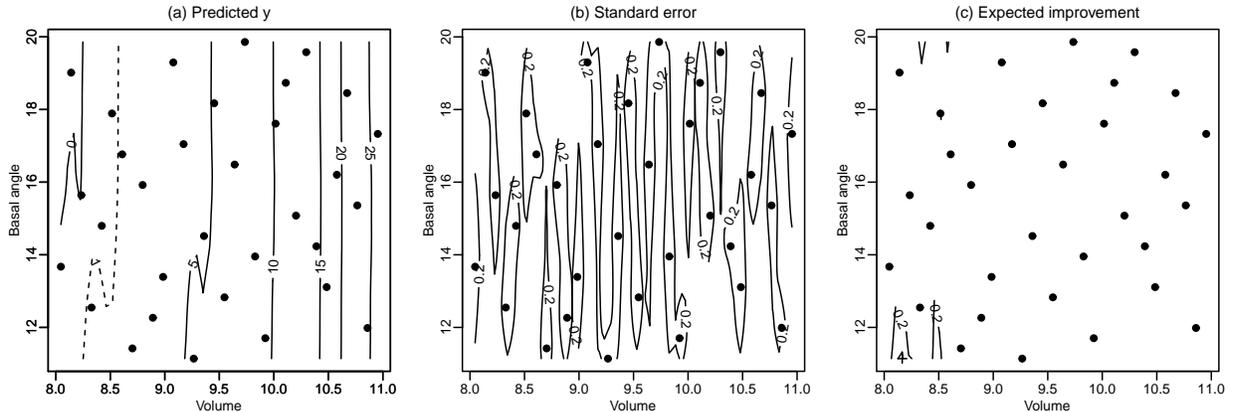,width=6.5in}}
\vspace{-0.5cm}
\caption{Analysis of the initial 32-run design for the volcano application:
(a) predicted height, $\hat{y}(\xVec)$, where $y = \sqrt{z}$;
(b) standard error, $s(\xVec)$; and
(c) expected improvement, $E[I(\xVec)]$.
The design points of the initial 32-run design are
shown as filled circles.
The new design point chosen by the EI criterion is shown as a ``$+$''
in the lower left corner of (c).
\label{fig:volcano}}
\end{figure}

There are 32 initial runs of the TITAN2D code.
They are located at the points shown in Figure~\ref{fig:volcano}(a).
The predicted flow height surface also shown in Figure~\ref{fig:volcano}(a)
relates to a GP model fit to the transformed simulator output $y = \sqrt{z}$.
This choice was made by trying GP models on three different scales:
the $z$ untransformed height;
$\log(z + 1)$, as chosen by \cite{BayBerCal2009};
and $\sqrt{z}$.
Our final choice of $y = \sqrt{z}$ results from inspection of
standard cross-validation diagnostics for GP models \citep{Jones1998}.

The dashed curve in Figure~\ref{fig:volcano}(a) shows the contour where $\hat{y}(\xVec) = 1$.
This maps out the contour of interest in the $(x_1, x_2)$ input space,
but it is based on predictions subject to error.
The standard errors in Figure~\ref{fig:volcano}(b) are substantial,
and sequential design via EI aims to improve the accuracy of the
estimate of the true $y(\xVec) = 1$ contour.

The EI criterion adapted for the contouring objective is defined in
Section~\ref{sect:EI:contour}.
It is computed for the initial design of the volcano example in
Figure~\ref{fig:volcano}(c).
EI suggests improving the accuracy of the contour
by taking the next run at $(x_1, x_2) = (8.2, 11.1)$.
Inspection of Figures~\ref{fig:volcano}(a) and~3(b) show that this location
is intuitively reasonable.
It is in the vicinity of the predicted $\hat{y}(\xVec) = 1$ contour and
has a relatively large predictive standard error.
Reducing substantial uncertainty in the vicinity of the estimated contour
is the dominant aspect of the EI measure of equation~(\ref{eq:EI_contour}).

The tidal-flow and volcano applications both have a 2D input space
for ease of exposition,
but the same approaches apply to higher dimensions,
where choosing runs in a sequential design would be more problematic
with ad hoc methods.

\section{Expected Improvement Criteria}\label{sect:EI}

In this section, we briefly define improvement and EI in general.
We then review two implementations,
specific to global optimization and contour estimation, respectively. 

Let $I(\xVec)$ be an improvement function defined for any $\xVec$
in the input space, $\chi$.
It depends on the scientific objective,
such as improving the largest $y$ found so far in maximization.
In general, it is formulated for efficient estimation of a
pre-specified computer-model feature, $\psi(y)$.
Typically, before taking another run,
$I(\xVec)$ is an unobserved function of $\xVec$,
the unknown computer-model output $y(\xVec)$,
the predictive distribution of $y(\xVec)$,
and the best estimate so far of $\psi(y)$.

Given a definition of $I(\xVec)$,
as its name suggests the corresponding EI criterion is given by
the expectation of $I(\xVec)$:
\begin{equation*}\label{eq:EI_general}
E[I(\xVec)] = \int I(\xVec) f(y | \xVec) \,  dy.
\end{equation*}
Here expectation is with respect to $f(y | \xVec)$,
the predictive distribution of $y(\xVec)$ conditional on all runs so far.
Assuming the sequential design scheme selects one new input point at a time,
the location of the new point, $\xnew$, is the global maximizer of $E[I(\xVec)]$
over $\xVec \in \chi$.

\subsection{EI for global optimization}\label{sect:EI:opt}% (local vs global)

%Much of the literature on EI in computer experiments has focussed
%on the estimation of the minimum output.

Finding the global minimum, $\psi(y) = \min\{y(\xVec): x \in \chi\}$,
of an expensive to evaluate function
is an extensively investigated optimization problem.
(Finding the maximum is reformulated as $\min -y(\xVec)$,
and the following results apply.)
\cite{Jones1998} proposed an efficient sequential solution
via the improvement function to assess the gain
if a new evaluation is made at $\xVec$.
The improvement function is
$$ I(\xVec) = \max\{\yminn{n} - y(\xVec), 0\}, $$
where $\yminn{n}$ is the minimum value of $y$ found so far with $n$ runs.
The objective is improved by $\yminn{n} - y(\xVec)$ if $\yminn{n} > y(\xVec)$,
otherwise there is no improvement.

The GP statistical model outlined in Section~\ref{sect:GP}
leads to a Gaussian predictive distribution for $f(y | \xVec)$,
i.e., $y(\xVec) \sim N(\hat{y}(\xVec), s^2(\xVec))$.
The Gaussian predictive model leads to a simple, closed form for the expected improvement:
\begin{equation}
E[I(\xVec)] = s(\xVec)\phi(u) + (\yminn{n}-\hat{y}(\xVec))\Phi(u),
\label{eq:EI:opt}
\end{equation}
where $u = (\yminn{n}-\hat{y}(\xVec))/s(\xVec)$,
and $\phi(\cdot)$ and $\Phi(\cdot)$ denote the standard normal
probability density function (pdf) and cumulative distribution function (cdf), respectively.

Large values of the first term support global exploration in regions of the input space
sparsely sampled so far, where $s(\xVec)$ is large.
The second term favours search where $\hat{y}(\xVec)$ is small,
which is often close to the location giving $\yminn{n}$,
i.e., local search.
This trade-off between local and global search
makes EI-based sequential design very efficient,
and it often requires relatively few computer-model evaluations
to achieve a desired accuracy in estimating $\min y$.

For instance, in the tidal-power application,
the EI surface in Figure~\ref{fig:tidal:ei20}(d)) indicates that the first follow-up run
is at the location giving the maximum predicted power (see Figure~\ref{fig:tidal:ei21}(a)).
Thus, the local-search component dominates.
Conversely, the suggested location for the second follow-up run
is in an unsampled region near the maximum predicted power (see Figure~\ref{fig:tidal:ei21}(c)).

Attempts have been made to control this local versus global trade-off
for faster convergence (that is, using as few runs as possible) to the true global minimum.
For instance, \cite{SchWelJon1998} proposed an
exponentiated improvement function, $I^g(\xVec)$, for $g \ge 1$.
With $g > 1$,
there is more weight on larger improvements
when expectation is taken to compute EI.
Such large improvements will have a non-trivial probability 
even if $\hat{y}(\xVec)$ is unfavourable, provided $s(\xVec)$ is sufficiently large. 
Hence, global exploration of high-uncertainty regions can receive more attention
with this adaptation.
Similarly,
\cite{Sobester2005} developed a weighted expected improvement function (WEIF)
by introducing a user-defined weight parameter $w \in [0, 1]$
in the \cite{Jones1998} EI criterion,
and \cite{Ponweiser2008} proposed clustered multiple generalized expected improvement.

\subsection{EI for contour estimation}\label{sect:EI:contour}%

\cite{Ranjan2008} developed an EI criterion
specific to estimating a threshold (or contour) of $y$.
They applied it to a 2-queue 1-server computer network simulator
that models the average delay in a queue for service.

Let the feature of interest $\psi(y)$ be the set of input vectors $\xVec$ 
defining the contour at level $a$:
\begin{equation}
S(a) = \{x: y(\xVec) = a\}.
\label{eqn:S}
\end{equation}
The improvement function proposed by \cite{Ranjan2008} is
$$ I(\xVec) = \epsilon^2(\xVec) - \min \{ (y(\xVec)-a)^2, \epsilon^2(\xVec) \}, $$
where $\epsilon(\xVec) = \alpha s(\xVec)$ for a positive constant $\alpha$
(e.g., $\alpha = 1.96$, corresponding to 95\% confidence/credibility under
approximate normality).
This improvement function defines a limited region of interest around $S(a)$
for further experimentation.
Point-wise, the extent of the region depends on the uncertainty $s(\xVec)$
and hence the tolerance $\epsilon(\xVec)$.

Under a normal predictive distribution, $y(\xVec) \sim N(\hat{y}(\xVec), s^2(\xVec))$,
the expectation of $I(\xVec)$  can again be written in closed form:
\begin{eqnarray}
E[I(\xVec)]
&=& [\epsilon^2(\xVec) -(\hat{y}(\xVec)-a)^2] \left(\Phi(u_2)-\Phi(u_1)\right) \nonumber \\
&+& s^2(\xVec) \left[ (u_2\phi (u_2)- u_1\phi(u_1)) - (\Phi(u_2)-\Phi(u_1))\right] \nonumber \\
&+& 2(\hat{y}(\xVec)-a)s(\xVec) \left(\phi (u_2)-\phi(u_1)\right),
\label{eq:EI_contour}
\end{eqnarray}
where $u_1 = (a - \hat{y}(\xVec) - \epsilon(\xVec))/s(\xVec)$ and
$u_2 = (a - \hat{y}(\xVec) + \epsilon(\xVec))/s(\xVec)$.
%and $\phi(\cdot)$ and $\Phi(\cdot)$
%be the standard normal pdf and cdf respectively.
% already defined
Like EI for optimization, the EI criterion in~(\ref{eq:EI_contour}) trades off
the twin aims of local search near the predicted contour of interest
and global exploration.
The first term on the right of~(\ref{eq:EI_contour}) recommends an input location
with a large $s(\xVec)$ in the vicinity of the predicted contour.
When it dominates, the follow-up point is often
essentially the maximizer of $\epsilon^2(\xVec) -(\hat{y}(\xVec)-a)^2$.
This consideration led to the new point in Figure~\ref{fig:volcano}(c) of
the volcano application, for instance.
The last term in~(\ref{eq:EI_contour}) gives weight to points
far away from the predicted contour with large uncertainties.
The second term is often dominated by the other two terms in the EI criterion.

The EI criterion in (\ref{eq:EI_contour}) can easily be extended to related aims.
For simultaneous estimation of $k$ contours $S(a_1),..., S(a_k)$,
with $S(\cdot)$ defined in~(\ref{eqn:S}),
the improvement function becomes
$$ I(\xVec) = \epsilon^2(\xVec)
- \min\left\{(y(\xVec)-a_1)^2,\ldots, (y(\xVec)-a_k)^2, \epsilon^2(\xVec) \right\}, $$
and the corresponding EI can also be written in a closed form.
% \citep{Bingham2009}. reference incomplete
When interest centres on the $100p$-th percentile, $\nu_p$, of the simulator output,
\cite{Roy2008} suggested sequential design to estimate $S(\nu_p)$ using
the improvement function
$$ I^g(\xVec) = \epsilon^g(\xVec) - \min\{ (y(\xVec) - \hat{\nu}_p)^g, \epsilon^g(\xVec)\}. $$
Here the contour of interest changes after every follow-up point
when $\hat{\nu}_p$ is estimated using Monte Carlo methods.
Note that $I^g(\xVec)$ for $g=2$ is the improvement function in (\ref{eq:EI_contour}).
Here $a=\hat{\nu}_p$ and hence not fixed throughout the sequential procedure. 
That is, $\hat{\nu}_p$ for choosing a run is likely to be different from one run to the next.
\cite{Bichon2009} adapted this criterion to estimate
the probability of rare events and system failure in reliability-based design optimization.

\section{Gaussian Process Models and Predictive Distributions}\label{sect:GP}

Evaluation of an EI criterion requires the computation of the expectation of $I(\xVec)$
with respect to the predictive distribution of $y(\xVec)$.
In principle, any predictive distribution can be used,
but for the method to be useful, 
it should faithfully reflect the data obtained up to the run in question.
In practice, treating the data from the computer-model runs
as a realization of a GP is nearly ubiquitous in computer experiments.
A GP model leads to a Gaussian predictive distribution,
which in turn leads to the closed form expressions
in (\ref{eq:EI:opt}) and (\ref{eq:EI_contour})
and easy interpretation of the trade off between local and global search.

A GP model is a computationally inexpensive statistical emulator of
a computer code.
A key feature of many codes is that they are deterministic:
re-running the computer model with the same values for all input variables
will give the same output values.
Such a deterministic function is placed within a statistical framework by
considering a given computer-model input-output relationship
as the realization of a stochastic process,
$Z(\xVec)$, indexed by the input vector.
A single realization of the process is non-random, hence the relevance for a deterministic
computer code.
For a continuous function, the process is usually assumed to be Gaussian,
possibly after transformation, as was done for the volcano application.

This GP or Gaussian Stochastic Process (GaSP) paradigm for modelling
a computer code dates back to
\cite{SacSchWel1989}, \cite{SacWelMit1989}, \cite{CurMitMor1991}, and \cite{Oha1992}.
Specifically, the code output function, $y(\xVec)$,
is treated as a realization of
$$ Y(\xVec) =  \mu(\xVec) + Z(\xVec), $$
where $\mu(\xVec)$ is a mean (regression) function in $\xVec$,
and $Z(\xVec)$ is a Gaussian process with mean $0$ and variance $\sigma^2$.

Crucial to this approach is the assumed correlation structure of $Z(\xVec)$.
For two configurations of the $d$-dimensional input vector,
$\xVec = (x_1,\ldots, x_d)$ and $\xVec' = (x'_1,\ldots, x'_d)$,
the correlation between $Z(\xVec)$ and $Z(\xVec')$ is denoted by $R(\xVec, \xVec')$.
Here, $R(\cdot, \cdot)$ is usually a parametric family of functions,
for which there are many choices \citep[e.g.,][Section 2.3]{SanWilNot2003}.
The computations for the applications in Section~\ref{sect:basic}
were based on a constant (intercept) regression only and a stationary
power-exponential correlation function,  
$$ R(\xVec, \xVec') = \exp\left(-\sum_{j=1}^d \theta_j | x_j - x'_j |^{p_j} \right). $$
Here, $\theta_j$ (with $\theta_j \geq 0$) and $p_j$ (with $1 \leq p_j \leq 2$) 
control the properties of the effect of input variable $j$ on the output. 
A larger value of $\theta_j$ implies greater sensitivity (activity) of $y$ 
with respect to $x_j$,
whereas a larger value of $p_j$ implies smoother behaviour of $y$ as a function of $x_j$.

Under this model the output values from $n$ runs of the code, $Y_1,\ldots,\allowbreak Y_n$,
have a joint multivariate normal distribution.
%$\mu({\bf X})=(\mu(x_1),..., \mu(x_n))'$
%and spatial covariance matrix $\Sigma$.
If the parameters in the statistical model---in the mean function, 
in the correlation function,
and $\sigma^2$---are treated as known,
the predictive distribution of $Y$ at a new $\xVec$ has a normal distribution:
$N(\hat{y}(\xVec), s^2(\xVec))$,
where $\hat{y}(\xVec)$ is the conditional mean of $Y(\xVec)$ given
$Y_1,\ldots,\allowbreak Y_n$,
and $s^2(\xVec)$ is the conditional variance.
Without assuming normality, $\hat{y}(\xVec)$ can also be interpreted as
the best linear unbiased predictor, and $s^2(\xVec)$ is the associated mean squared error.
In practice, the unknown parameters have to be estimated,
usually by maximum likelihood or Bayesian methods.
The predictive distribution is then only approximately normal.
Moreover, Bayesian estimation of the correlation parameters may be necessary
to capture all sources of uncertainty in the predictive distribution.

As mentioned already,
neither a GP model nor a normal predictive distribution are essential
for sequential design with an EI criterion.
%One might consider another statistical metalled for emulating the simulator outputs.
For instance, \cite{Chipman2012} used the optimization improvement function of \cite{Jones1998}
with Bayesian additive regression trees (BART).
Thus, the emulator was a non-parametric ensemble of tree models.

\section{Other EI-based criteria}\label{sect:other}%Noisy response, robust design

Over the last two decades, a plethora of EI-based criteria have been proposed for
other scientific and engineering objectives.
%We now outline various extensions of EI for other scientific and engineering objectives.
%however, EI criteria for estimating other process features like contours, percentiles and probability %of system failure, have also been developed.

Applications can involve several outputs of interest.
For instance,
constrained optimization problems arise 
where the code generating the objective function $y(\xVec)$ or another code 
gives values for a constraint function, $c(\xVec)$, (or several functions).
For a feasible solution, $c(\xVec)$ must lie in $[a, b]$.
If $c(\xVec)$ is also expensive to compute, 
one can build an emulator, $\hat{c}(\xVec)$, for it too.
The predictive distribution for $c(\xVec)$ leads to an estimate of the 
probability that $a < c(\xVec) < b$ for any new run $\xVec$ under consideration.
EI in~(\ref{eq:EI:opt}) is multiplied by this probability of feasibility
to steer the search to locations where EI for the objective $y(\xVec)$ 
is large and $c(\xVec)$ is likely to be feasible \citep{SchWelJon1998}.
For a code with multivariate output,
\cite{Henkenjohann2007} proposed an EI criterion for estimating
the global maximum of the desirability scores of simulator outputs.

\cite{lehman_etal} developed improvement functions for finding $M$- and $V$-robust designs
for optimization of an engineering process.
Here the simulator inputs include
both controllable and environmental (uncontrollable noise) variables.
(``Uncontrollable'' here means in the field; all inputs are typically set
at specified values in a simulator run.)
For a given configuration of the control variables, $\xVec_c$,
let $\mu(\xVec_c)$ and $\sigma^2(\xVec_c)$ be the unknown mean and variance
of the simulator output $y$ with respect to the distribution of the environmental variables.
An $M$-robust engineering design minimizes $\mu(\xVec_c)$ with respect to $\xVec_c$
subject to a constraint on $\sigma^2(x_c)$,
whereas  $V$-robust engineering design minimizes $\sigma^2(x_c)$ subject
to a constraint on $\mu(x_c)$.

The inclusion of measurement error
(or equivalently, considering a nondeterministic simulator)
is becoming more popular in computer experiments,
often due to unavoidable simulator biases and inaccurate modelling assumptions.
Minimizing a noisy mean output response is perhaps undesirable,
and \cite{Ranjan2013} recommended minimizing a lower quantile, $q(\xVec)$,
via an estimate $\hat{q}(\xVec)$
from the predictive distribution, e.g.,
$\hat{q}(\xVec) = \hat{y}(\xVec) - 1.96 s(\xVec)$ under a normal predictive distribution.
The proposed improvement function is
$I(\xVec) =\max\{0, \hat{q}_{min}^{(n)} - q(\xVec)\}$,
where $\hat{q}_{min}^{(n)}$ is the minimum $\hat{q}(\xVec)$ from $n$ runs so far,
and $q(\xVec) = y(\xVec) - 1.96 s(\xVec)$ is an unobservable random quantity.
Treating $s(\xVec)$ as non-stochastic and assuming $y(\xVec)\sim N(\hat{y}(\xVec), s^2(\xVec))$,
the corresponding EI criterion is
\begin{equation}\label{eq:EI_noisy}
E[I(\xVec)] = s(\xVec) \phi(u) + (\hat{q}_{min}^{(n)} - \hat{y}(\xVec) + 1.96s(\xVec))\Phi(u),
\end{equation}
where $u = (\hat{q}_{min}^{(n)} - \hat{y}(\xVec) + 1.96 s(\xVec))/ s(\xVec)$.
Like the EI criterion in (\ref{eq:EI:opt}),
EI in (\ref{eq:EI_noisy}) facilitates the trade-off between local and global search.
One can easily generalize this EI criterion to $E[I^g(\xVec)]$ \citep[as in][]{SchWelJon1998}
or introducing a user specified weight \citep[as in][]{Sobester2005}.

For complex physical phenomena like climate and tidal power,
multiple computer simulators with different computational demands 
are often available for experimentation.
For instance, 
there are 2D and 3D codes for the tidal-power application; 
the 3D version is a higher-fidelity representation of reality 
but is much more expensive to run.
They can be combined to obtain more informed prediction,
% as compared to that obtained from one simulator data.
and \cite{Huang2006} proposed augmented expected improvement
for finding the global minimum of the highest-fidelity process, subject to noise.

%If a sequential strategy with batches of runs at a time are more convenient,
%\cite{SchWelJon1998} proposed a generalized expected improvement criterion.

%Alternatively, one can minimize the \emph{integrated expected improvement criterion} (in the spirit of %integrated mean squared error by \cite{SacWelMit1989}) for choosing a batch of points.

\section{Summary}

The essence of these approaches for sequential computer experiments is to formulate
the scientific objective through an improvement function.
Following some initial runs, 
the next run is chosen to maximize the expected improvement.
In contrast to physical experiments,
sequential design is convenient, with the computer handling
the logistics of iterating analysis of the data so far, choice of the next run,
and making the new run.

With objectives like optimization and contouring in high-dimensional applications,
sequential strategies are efficient in terms of solving the problem with a
relatively small number of runs.
For these reasons, we expect this area of the design of computer experiments
will continue to receive considerable research attention from methodologists and users.

Of course, the usefulness of this strategy depends 
on having a computer model that provides a satisfactory 
description of the physical process of interest. 
Such models have to be checked by reference to real data from the physical process
\citep{BayBerPau2007}. 
However, once a model has been adequately validated it provides an efficient route 
to achieving the objectives discussed here.

%----------------------------------------------------------------------%
%\newpage
\bibliographystyle{agsm}
\bibliography{SSCchap}
%----------------------------------------------------------------------%

\end{document}